# Comparison of ARIMA, ETS, NNAR and hybrid models to forecast the second wave of COVID-19 hospitalizations in Italy

Gaetano Perone[†]

[†]*Department of Management, Economics and Quantitative Methods, University of Bergamo, Italy*

**Abstract**

Coronavirus disease (COVID-19) is a severe ongoing novel pandemic that has emerged in Wuhan, China, in December 2019. As of October 13, the outbreak has spread rapidly across the world, affecting over 38 million people, and causing over 1 million deaths. In this article, I analysed several time series forecasting methods to predict the spread of COVID-19 second wave in Italy, over the period after October 13, 2020. I used an autoregressive model (ARIMA), an exponential smoothing state space model (ETS), a neural network autoregression model (NNAR), and the following hybrid combinations of them: ARIMA-ETS, ARIMA-NNAR, ETS-NNAR, and ARIMA-ETS-NNAR. About the data, I forecasted the number of patients hospitalized with mild symptoms, and in intensive care units (ICU). The data refer to the period February 21, 2020–October 13, 2020 and are extracted from the website of the Italian Ministry of Health (www.salute.gov.it). The results show that i) the hybrid models, except for ARIMA-ETS, are better at capturing the linear and non-linear epidemic patterns, by outperforming the respective single models; and ii) the number of COVID-19-related hospitalized with mild symptoms and in ICU will rapidly increase in the next weeks, by reaching the peak in about 50-60 days, i.e. in mid-December 2020, at least. To tackle the upcoming COVID-19 second wave, on one hand, it is necessary to hire healthcare workers and implement sufficient hospital facilities, protective equipment, and ordinary and intensive care beds; and on the other hand, it may be useful to enhance social distancing by improving public transport and adopting the double-shifts schooling system, for example.

**Keywords**: COVID-19; outbreak; second wave; Italy; hybrid forecasting models; ARIMA; ETS; NNAR.

**JEL Classification**: C22; C53; I18.

e-mail: gaetano.perone@unibg.it.

# 1. Introduction

Coronavirus disease (COVID-19) is a severe ongoing novel pandemic that has officially emerged in Wuhan, China, in December 2019. As of October 13, 2020, it has affected 215 countries and territories, with over 38 million cases and approximately 1.1 million deaths (Worldometer, 2020). At the time of writing, the most affected countries are both advanced and developing countries, such as Argentina, Brazil, Colombia, France, India, Peru, Russia, Spain, and the USA. In the last two weeks, several European countries, including Italy, saw a worrying surge of COVID-19 infections.

Italy was the first European country to be severely hit by COVID-19, and it has been one of the main epicenters of the pandemic for about two months, i.e. from mid-February 2020 to mid-April 2020, when the outbreak reached the first peak. Then, the epidemic curve progressively decreased until mid-August 2020, and after that the spread of infection re-accelerated again until today. As of October 13, Italy has suffered 36,246 deaths and 365,467 cases.

The likelihood of a second wave is real and makes it necessary to predict the future epidemic evolution, in plan to buy medical devices and healthcare facilities, and to manage health centers, clinics, hospitals, and ordinary and intensive care beds.

Thus, the major goal of this paper is to provide short and mid-term forecasts of the patients hospitalized from COVID-19 over the period after October 13, 2020. In fact, the COVID-19-related hospitalizations trends allow to have a clear picture of the overall stress and pressure on the national health care system. In particular, I implemented and compared three different time series forecast techniques and their feasible hybrid combinations: autoregressive moving average (ARIMA) model, innovations state space models for exponential smoothing (ETS), neural network autoregression (NNAR) model, ARIMA-ETS model, ARIMA-NNAR model, ETS-NNAR model, and ARIMA-ETS-NNAR model.

I organized the rest of this paper as follows. In section 2, I quickly discussed the relevant literature. In section 3, I presented the data used in the analysis and discussed the empirical strategy. In section 4, I discussed the main findings and policy implications. Finally, in section 5, I provided some conclusive considerations.

# 2. Related literature

From the beginning of the 2020, an increasing body of literature has attempted to forecast the spread of the COVID-2019 outbreak using different approaches (Bhardwaj, 2020; Fanelli and Piazza, 2020; Giordano et al., 2020; Nesteruk. 2020; Tuite et al., 2020; Wu et al., 2020; Xu et al., 2020; Zhao et al., 2020; Zhou et al., 2020). The most used are ARIMA (Alzahrani et al. 2020; Benvenuto et al., 2020, Ceylan, 2020; Perone, 2020a & b), ETS (Bhandary et al., 2020; Cao et al., 2020; Joseph et al., 2020), artificial neural network models (Melin et al., 2020; Wieczorek et al., 2020), models derived from the susceptible-infected-removed (SIR) basic approach (Fanelli and Piazza, 2020; Giordano et al., 2020; Nesteruk, 2020; Wu et al., 2020; Zhou et al., 2020), and hybrid models (Chakraborty and Ghosh, 2020; Hasan et al., 2020; Singh et al., 2020; Swaraj et al., 2020).

The implementation and the comparison of them, except for SIR models, represent the core of this paper. Thus, in table 1, I reported 18 international studies that used single or hybrid ARIMA, ETS, and neural network to forecast the transmission patterns of COVID-19 across the world.

Table 1. Eighteen selected studies on COVID-19 forecasts, which use a single or hybrid ARIMA, ETS, and/or neural network approach.

| Authors | Data used | Method | Investigated area |
|---|---|---|---|
| Alzahrani et al. (2020) | Confirmed | ARIMA | Saudi Arabia |
| Aslam (2020) | Confirmed, Active, recovered, & deceased | KF-ARIMA, HW, & SutteARIMA | Pakistan |
| Bhandary et al. (2020) | Confirmed | ARIMA, ETS, & SIR | Nepal |
| Cao et al. (2020) | Confirmed | ARIMA, ARIMAX, ETS, & SEIQDR | China |
| Ceylan (2020) | Confirmed | ARIMA | France, Italy, & Spain |
| Chakraborty and Ghosh (2020) | Confirmed | ARIMA-WBF | Canada, France, India, & South Korea |
| Hasan (2020) | Confirmed | ANN-EEMD MA, & REG | World (aggregate) |
| Joseph et al. (2020) | Confirmed | ETS & INGARCH | Nine countries |
| Khan & Gupta (2020) | Confirmed | ARIMA & NNAR | India |
| Kufel (2020) | Confirmed | ARIMA | Several European countries |
| Melin et al. (2020) | Confirmed | ME-ANN | Mexico |
| Perone (2020a, 2020b) | Confirmed, & deceased | ARIMA | Italy, Russia, & USA |
| Ribeiro et al. (2020) | Confirmed | ARIMA, CUBIST, RF, RIDGE, SVR, & SEL | Brazil |
| Singh S. et al. (2020) | Deceased | ARIMA-WBF | France, Italy, Spain, UK, & USA |
| Swaraj et al. (2020) | Confirmed | ARIMA, NNAR, & ARIMA-NNAR | India |
| Wieczorek et al. (2020) | Confirmed | ANN | Several countries & regions |
| Yonar H. et al. (2020) | Confirmed | ARIMA & B/W LES | G8 countries |

Notes: ANN, artificial neural network; ARIMA, autoregressive integrated moving average; ARIMAX, ARIMA with exogenous variables; B/W LES, Brown/Holt linear exponential smoothing method; CUBIST, cubist regression ; EEMD, ensemble empirical model decomposition; HW Holt-Winters method; INGARCH, integer-valued generalized autoregressive conditional heteroskedastic; KF, Kalman filter; MA, moving average; ME-ANN, multiple ensemble artificial neural network; NNAR, nonlinear autoregressive neural network; REG, linear regression; RF, random forest; RIDGE, ridge regression; SEIQDR, susceptible-infected but undetected-infected quarantined-suspected-discharged; SEL, stacking-ensemble learning; SIR, susceptible-infectious-recovered; SutteARIMA, α-Sutte Indicator and ARIMA; SVR, support vector regression; WBF, Wavelet-bases forecasting.

## 3. Materials and methods

The data used in this article refer to the real-time number of COVID-19 patients with mild symptoms and in ICU in Italy from February 21, 2020 to October 13, 2020, for 236 observations. I extracted the data from the official Italian Ministry of Health's website (www.salute.gov). The confirmed COVID-19-related hospitalizations trends are shown in Figure 1.

Data show that both patients hospitalized with mild symptoms and in ICU reached a first peak on April 4, 2020; after that, they followed a downward trend until mid-August, before re-accelerating again from the end of September 2020 to mid-October 2020. I compute the forecasts by using different statistical techniques and their combination. Specifically, I use a univariate linear autoregressive integrated moving average (ARIMA) models, univariate exponential smoothing state space model (ETS), univariate linear and nonlinear neural network autoregression (NNAR) models, hybrid ARIMA-ETS model, hybrid ARIMA-NNAR model, and hybrid ETS-NNAR model.

The combination of different times series forecast methods should allow to maximize the chance of capturing both the linear (if any) and nonlinear epidemic patterns (Zhang, 2003; Pannigrahi and Behera, 2017), and it is useful with phenomena like COVID-19 epidemic, which seems characterized by linear and nonlinear dynamics and components (Batista, 2020, Gupta and Pal, 2020). As well established from the seminal work of Bates and Granger (1969), combining techniques with unique properties may allow to achieve better performance and forecast accuracy.[2] The models are calculated as follows:

- ARIMA models are detected by applying the "auto.arima" function included in the package "forecast" (in R environment) and developed by Hyndman and Khandakar (2008).[3] This function follows sequential steps to identify the best ARIMA models, i.e. the number of p parameters of the autoregressive process (AR), the order i of differencing (I), and the number of q parameters of the moving average process (MA).[4] It combines unit root tests,[5] and the minimization of the following estimation methods: the bias-corrected Akaike's information criterion (AICc),[6] and the maximum likelihood estimation (MLE). The unit root tests allow to identify the order of differencing; while the AICc and the MLE methods allow to identify the optimal parameters of the AR and MA processes. Finally, the overall goodness of fit is tested by using five common forecast accuracy measures: mean absolute error (MAE), mean absolute percentage error (MAPE), mean absolute scaled error (MASE), and root mean square error (RMSE);

---

[2] See also, for example, Fallah et al. (2018).
[3] A description of the "auto.arima" function is provided by Hyndman and Athanasopoulos (2018, section 8.7).
[4] This definition refers to non-seasonal ARIMA model.
[5] I use both the augmented Dickey-Fuller's test (ADF) (1981) and the Kwiatkowsky, Phillips, Schmidt, and Shin's test (KPSS) (1992). In fact, as stated by Gujarati and Porter (2009), there is not a recognized uniformly powerful test for detecting unit root.
[6] The AICc is a bias-corrected version of the original Akaike information criterion (AIC), proposed by Sugiura (1978) and Hurvich and Tsai (1989), which performs significantly better than the latter in both small and moderate sample sizes, as in this case (Hurvich and Tsai, 1989).

- ETS models are identified by using the "ets" function included in the package "forecast" (in R environment), developed by Hyndman et al. (2008).[7] ETS simple models comprise two main equations: a forecast equation and a smoothing equation. By implementing the last two equations into an innovation state space model, it is possible to get an observation/measurement equation and a transition/state equation, respectively. The first equation allows to describe the observed data; while the second equation allows to describe the behavior of the unobserved states. The states refer to the level, trend, and seasonality. In particular, I use the AICc metric for choosing the best ETS model. The goodness of fit is tested by using MAE, MAPE, MASE, and RMSE metrics;

- NNAR models are identified by using the "nnetar" function included in the package "caret" (R environment), developed by Hyndman, O'Hara, and Wang.[8] For non-seasonal data, I can describe NNAR models with the notation NNAR (p,k), where p denotes the number of non-seasonal lags used as inputs, and k means the number of nodes/neurons in the hidden layer. NNAR (p,k) is the same as AR process, but with nonlinear functions. The optimal number of non-seasonal lags is obtained by using the AICc metric, and the optimal number of neurons is identified by calculating (p+P+1)/2, where p is the non-seasonal AR order, and P is the seasonal AR order (if any). Finally, the goodness of fit is investigated using MAE, MAPE, MASE, and RMSE metrics;

- Hybrid models are identified by using the "hybridModel" function included in the package "forecastHybrid" (in R environment), developed by Shaub and Ellis.[9] To combine the single time series forecasting methods, I proceed as follows: i) first, I apply the Box-Cox (1964) power transformation to the inputs to make normality assumption more plausible; and ii) then, I implement the cross validation errors ("cv.errors"), which allow to give more weight to the models that perform relatively better, by producing the best forecast; and iii) finally, MAE, MAPE, RMSE, and Theil's U are used for validating the weighting procedure.

The estimated basic equation for ARIMA model is the following (Davidson, 2000):

$$\Delta^d y_t = \phi_1 \Delta^d y_{t-1} + \cdots \phi_p \Delta^d y_{t-p} + y_1 \varepsilon_{t-1} + \cdots y_q \varepsilon_{t-q} + \varepsilon_t \qquad [1]$$

---

[7] A description of the "ets" function is provided by Hyndman and Athanasopoulos (2018, section 7.6)
[8] A description of the "nnetar" function is provided by Hyndman and Athanasopoulos (2018, section 11.3).
[9] A detailed description of the "forecastHybrid" function is provided at the ULR: https://cran.r-project.org/web/packages/forecastHybrid/forecastHybrid.pdf

Where $\Delta^d$ the second difference operator, $p$ is the lag order of the AR process, $\phi$ is the coefficient of each parameter $p$, $q$ is the order of the MA process, $\gamma$ is the coefficient of each parameter $q$, and $\varepsilon_t$ denotes the residuals of errors in time $t$.

The estimated equations for basic ETS (A,N,N) model with additive error is the following (Hyndman and Athanasopoulos, 2018, section 7)[10]:

Forecast equation: $\hat{y}_{t+1|t} = l_t$ [2]

Smoothing equation: $l_t = l_{t-1} + \alpha(y_t - l_{t-1})$ [3]

Where $l_t$ is the new estimated level, $\hat{y}_{t+1|t}$ denotes each one-step-ahead prediction for time $t+1$, which results from the weighted average of all the observed data, $0 \leq \alpha \leq 1$ is the smoothing parameter which controls the rate of decrease of the weights, and $y_t - l_{t-1}$ is the error at time $t$. So, each forecasted observation is the sum of the previous level and an error. For each type of errors, additive or multiplicative, there is a specific probability distribution. For a model with additive errors, such in this case, it is assumed that errors follow a normal distribution. Thus, the equations [2] and [3] can be rewritten as follows, respectively:

Observation equation: $y_t = l_{t-1} + \varepsilon_t$ [4]

Transition equation: $l_t = l_{t-1} + \alpha\varepsilon_t$ [5]

The equations [4] and [5] represent the innovations state space models that underlie the exponential smoothing methods.

The basic form of the neural network autoregression equation is the following (Hyndman and Athanasopoulos 2018, section 11.3):

$y_t = f(y_{t-1}) + \varepsilon_t$ [6]

Where $y_{t-1} = (y_{t-1}, y_{t-2} ..., y_{t-n})'$ is a vector containing the lagged values of the observed data, $f$ is the neural network with $n$ hidden neurons in a single layer, and $\varepsilon_t$ is the error. Finally, the hybrid models are simple a combination of the described single models.

---

[10] How stated by Hyndman and Athanasopoulos (2018, section 7), the innovations state space models for exponential smoothing encompass 15 different methods. For each method there are two models: one with additive errors and one with multiplicative errors, for a total of 30 models. The equation [2] refers to the simplest of the ETS models.

Figure 1. Patients hospitalized with mild symptoms and in intensive care units, from February 21, 2020 to October 13, 2020.

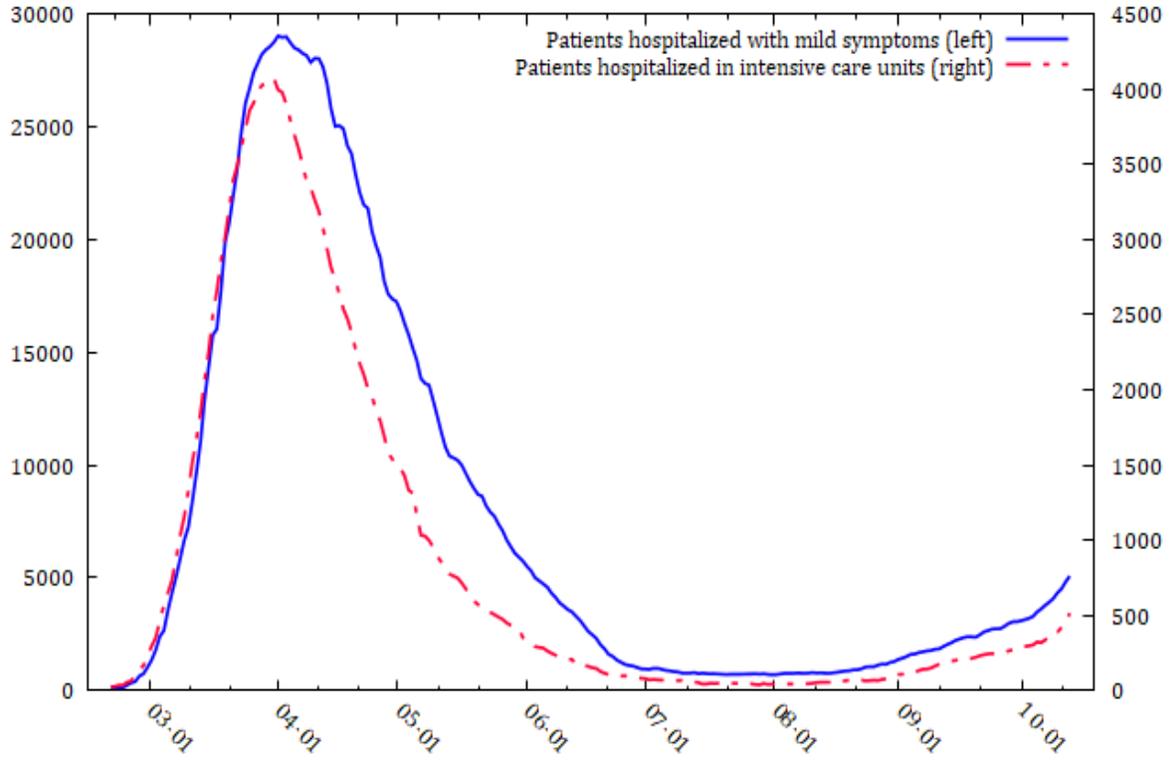

Source: Italian Ministry of Health (www.salute.gov.it).

## 4. Results and discussion

In tables 2, 3, 4 and 5, I reported the main forecast accuracy measures, and the best selected parameters for the single and hybrid models. For the patients hospitalized with mild symptoms (Tables 2 and 3), the optimal models are: ARIMA (4,2,4), ETS (A,Ad,N),[11] NNAR (4,2), hybrid ARIMA (4,2,4)-NNAR (7,4), hybrid ARIMA (4,2,4)-ETS (A,Ad,N), hybrid ETS (A,Ad,N)-NNAR (7,4), and hybrid ARIMA (4,2,4)-ETS(A,Ad,N)-NNAR (7,4). For the patients hospitalized in ICU (Tables 4 and 5), the optimal single and hybrid models are: univariate ARIMA (3,2,7), univariate ETS (A,A,N), NNAR (4,2), hybrid ARIMA(4,2,3)-NNAR(6,4), hybrid ARIMA(4,2,3)-ETS(A,A,N),[12] hybrid ETS(A,A,N)-NNAR(6,4), and hybrid ARIMA(4,2,3)-ETS(A,A,N)-NNAR(6,4).[13]

According to Lewis' interpretation (Lewis 1982, p. 40), since MAPE is always lower than 10, all the predictive models are highly accurate. In particular, the forecasts show a forecast accuracy (100-MAPE) ranging from 95.68% to 97.84%. Since MASE and Theil's U are always and significantly lower than 1, all the proposed forecasting models performed better than the forecasts from the (no-change) "naïve" methods, i.e. the forecasts with no adjustments for casual factors (Hyndman and Koehler, 2006). This allows to justify the use of more complex and sophisticated models, such as ARIMA, ETS, NNAR, and their hybrid combination.

---

[11] The selected ETS model is also known as damped trend method with additive errors.
[12] The selected ETS model is also known as Holt's linear trend method.
[13] The coefficients of the parameters estimated for the single and hybrid ARIMA and ETS models are provided in Tables A1, A2, A3, & A4 (Appendix A).

In tables 6 and 7, I compared the single and hybrid models by considering the minimization of MAE, MAPE, and RMSE. For the patients hospitalized with mild symptoms, hybrid ARIMA-ETS is better than the respective single models in 2 measures out of 6; the hybrid ARIMA-NNAR and ETS-NNAR are better than the respective single models in all the accuracy measures; and the hybrid ARIMA-ETS-NNAR is better than respective single models in 8 measures out of 9. For the patients hospitalized in ICU, hybrid ARIMA-ETS is better than the single models in 3 measures out of 6; and hybrid ARIMA-NNAR is better than the respective single models in all the accuracy measures; ETS-NNAR is better than the respective single models in 5 measures out of 6; and ARIMA-ETS-NNAR is better than respective single models in 8 measures out of 9.

The most reliable and accurate single and hybrid models are ARIMA and ARIMA-NNAR, respectively. The results show that hybrid models always outperformed the single models, with the only exception of ARIMA-ETS. The best models are ARIMA-NNAR, ETS-NNAR, and ARIMA-ETS-NNAR.

In table 8, I also compute the percentage efficiency gains – in terms of MAE and RMSE minimization – from using the three best hybrid models. Specifically, hybrid ARIMA-NNAR outperforms single ARIMA and NNAR models from 1.87% to 8.91% on MAE, and from 4.19% to 9.47% on RMSE. Hybrid ETS-NNAR outperforms single ETS and NNAR models from 3.01% to 9.02% on MAE, and from 2.68% to 12.4% on RMSE. Finally, hybrid ARIMA-ETS-NNAR outperforms single ARIMA, ETS, and NNAR models from 0.69% to 9.61% on MAE, and from 0.1% to 12.4% on RMSE.

In figures 2 to 8, I fit all the seven models for both the time series. The light blue area shows the prediction interval at 80%, and the dark blue area shows the prediction interval at 95%. The forecasts of the best models show that the number of patients hospitalized with mild symptoms and in ICU will significantly increase in the next 45 days, i.e. from October 14, 2020 to November 27, 2020. However, if the number of patients hospitalized with mild symptom will approximately stabilize between the end of November 2020 and the beginning of December 2020, the number of patients in ICU will require more time to reach the *plateau*. This is also confirmed by the other forecasting methods, except of ETS.

ARIMA-NNAR, ETS-NNAR, and ARIMA-ETS-NNAR show that: i) after 10 days (October 23), the number of patients hospitalized with mild symptoms will be 8,829, 8,032, or 8,125, respectively; ii) after 20 days (November 2), they will be 14,038, 12,058, or 11,877; and iii) after 45 days (November 27), they will be 20,795, 14,823 or 16,052. About the number of patients hospitalized in ICU, ARIMA-NNAR, ETS-NNAR, and ARIMA-ETS-NNAR show that: i) after 10 days, the required intensive care beds will be 1,177, 1,040 or 1,100; ii) after 20 days, they will be 2,065, 1,446, or 1,829; and iii) after 45 days, they will be 3,320, 1,907 or 3,030.[14]

---

[14] The forecasted values are shown in Table A5 (Appendix A).

Thus, a second wave of COVID-19 is expected in the next two months, with a peak in mid-December 2020. This has several policy implications both for the national health care system and economic activities. In particular, the predictions seem to stress the importance of implementing adequate containment measures and an increasing number of ordinary and intensive care beds, of hiring further healthcare personnel, and of buying care facilities, protective equipment, and ventilators to fight the infection and reduce deaths.[15]

However, the opportunity of implementing more or less restrictive non-pharmaceutical interventions (NIP) to tackle the epidemic – such as social distancing, travel ban, the use of face mask, hand hygiene, and bar and restaurant restrictions (ECDC, 2020) – should be carefully evaluated because of the negative impact on the overall economic activity. In fact, according to Fitch Rating's (2020) previsions, the first wave of COVID-19 and the consequent massive lockdown measures could have already caused up to 9.5% contraction for the Italian 2020 GDP.

Thus, to avoid new strict lockdown measures and further economic loss, it seems important to ensure the strict compliance with the basic COVID-19 control measures, such as social distancing, hand hygiene, and the use of protective equipment, rather than restricting or even closing public and private economic activities, such as provided by the recent Italian Prime Minister's Decrees of October 13, 2020. In fact, if the extension of the national state of emergency is well justified, the hours restriction for bakery, bar, ice cream shops, pubs, restaurant, and retail trade activities, and the total closure of all indoor and outdoor dance halls, discos, and similar spaces may seriously worsen the current Italy's economic downturn.[16]

By the contrary, it may be more useful to expand public transport, adopt the double-shifts schooling system, and isolate older and vulnerable people. Finally, a better balance between public health, public utility, and freedom of economic initiative is recommended.

Table 2. Forecast accuracy measures of the single and hybrid models for patients hospitalized with mild symptoms.

| Model | MAE | MAPE | MASE | RMSE | ACF1 | Theil's U |
|---|---|---|---|---|---|---|
| ARIMA | 112.2358 | 2.6081 | 0.4218 | 201.0097 | 0.0207 | - |
| ETS | 121.3297 | 2.5805 | 0.456 | 219.229 | 0.1191 | - |
| NNAR | 116.8677 | 2.2939 | 0.4392 | 210.4792 | -0.0071 | - |
| ARIMA-ETS | 113.9646 | 3.4691 | - | 206.49 | 0.0495 | 0.9006 |
| ARIMA-NNAR | 106.4554 | 2.1901 | - | 190.542 | -0.0238 | 0.3536 |
| ETS-NNAR | 113.8641 | 2.1585 | - | 204.9186 | 0.0119 | 0.3542 |
| ARIMA-ETS-NNAR | 111.4661 | 2.16 | - | 201.0401 | 0.0178 | 0.3621 |

---

[15] This is consistent with my recent paper (Perone, 2021), in which I showed that the Italian health care system saturation played a key role in explaining the variability of COVID-19 mortality. In particular, ordinary and intensive care beds saturation allowed to explain almost up to 90% of the COVID-19 mortality, at the first peak of the epidemic.
[16] Further details about the restrictions and control measures implemented by the Italian government (on October 13, 2020) are available at: https://www.gazzettaufficiale.it/eli/id/2020/10/13/20A05563/sg.

Table 3. Structure of the single and hybrid models for patients hospitalized with mild symptoms.

| Model | Components | AICc | $\sigma^2$ | Structure | Weight |
|---|---|---|---|---|---|
| ARIMA | ARIMA | 3,169.07 | 40,750 | (4,2,4) | - |
| ETS | ETS | 1,633.72 | 2.0419 | (A,Ad,N) | - |
| NNAR | NNAR | - | 0.7027 | (4,2) | - |
| ARIMA-ETS | ARIMA | 3,169,07 | 42,193 | (4,2,4) | 0.46 |
|  | ETS | 3,846.92 | 222.0379 | (A,Ad,N) | 0.54 |
| ARIMA-NNAR | ARIMA | 3,169,07 | 42,193 | (4,2,4) | 0.698 |
|  | NNAR | . | 34,249 | (7,4) | 0.302 |
| ETS-NNAR | ETS | 3,846,92 | 222.0388 | (A,Ad,N) | 0.73 |
|  | NNAR | - | 31,658 | (7,4) | 0.27 |
| ARIMA-ETS-NNAR | ARIMA | 3,169.07 | 42,193 | (4,2,4) | 0.387 |
|  | ETS | 3,846.92 | 222.0388 | (A,Ad,N) | 0.455 |
|  | NNAR | - | 31,371 | (7,4) | 0.158 |

Table 4. Forecast accuracy measures of the single and hybrid models for patients hospitalized in ICU.

| Model | MAE | MAPE | MASE | RMSE | ACF1 | Theil's U |
|---|---|---|---|---|---|---|
| ARIMA | 12.2536 | 3.4997 | 0.3327 | 20.4164 | -0.0071 | - |
| ETS | 13.609 | 3.4292 | 0.3696 | 23.2846 | 0.2078 | - |
| NNAR | 12.7657 | 3.2256 | 0.3467 | 20.9578 | -0.1716 | - |
| ARIMA-ETS | 12.6339 | 3.4923 | - | 21.3438 | -0.0258 | 0.6753 |
| ARIMA-NNAR | 12.0248 | 3.2221 | - | 19.5603 | -0.0633 | 0.5778 |
| ETS-NNAR | 12.3817 | 3.229 | - | 20.3967 | -0.0623 | 0.5851 |
| ARIMA-ETS-NNAR | 12.3018 | 3.2007 | - | 20.3963 | -0.0594 | 0.5785 |

Table 5. Structure of the single and hybrid models for patients hospitalized in ICU.

| Model | Components | AICc | $\sigma^2$ | Structure | Weight |
|---|---|---|---|---|---|
| ARIMA | ARIMA | 2,103.88 | 420.4 | (3,2,7) | - |
| ETS | ETS | 1,154.31 | 0.7412 | (A,A,N) | - |
| NNAR | NNAR | - | 0.0065 | (4,2) | - |
| ARIMA-ETS | ARIMA | 2,106.15 | 450.4 | (4,2,3) | 0.487 |
|  | ETS | 2,775.88 | 23.0116 | (A,A,N) | 0.513 |
| ARIMA-NNAR | ARIMA | 2,106.15 | 450.4 | (4,2,3) | 0.611 |
|  | NNAR | - | 360.5 | (6,4) | 0.389 |
| ETS-NNAR | ETS | 2,775.88 | 23.0116 | (A,A,N) | 0.621 |
|  | NNAR | - | 344.7 | (6,4) | 0.379 |
| ARIMA-ETS-NNAR | ARIMA | 2,106.15 | 450.4 | (4,2,3) | 0.367 |
|  | ETS | 2,775.81 | 23.0116 | (A,A,N) | 0.386 |
|  | NNAR | - | 333.6 | (6,4) | 0.248 |

Table 6. Comparison between hybrid and single models for patients hospitalized with mild symptoms.

| Hybrid models | Single models | MAE | MAPE | RMSE |
|---|---|---|---|---|
| ARIMA-ETS | ARIMA | Single | Single | Single |
|  | ETS | Hybrid | Single | Hybrid |
| ARIMA-NNAR | ARIMA | Hybrid | Hybrid | Hybrid |
|  | NNAR | Hybrid | Hybrid | Hybrid |
| ETS-NNAR | ETS | Hybrid | Hybrid | Hybrid |
|  | NNAR | Hybrid | Hybrid | Hybrid |
| ARIMA-ETS-NNAR | ARIMA | Hybrid | Hybrid | Hybrid |
|  | ETS | Hybrid | Hybrid | Hybrid |
|  | NNAR | Hybrid | Hybrid | Single |

Table 7. Comparison between hybrid and single models for patients hospitalized in ICU.

| Hybrid model | Single model | MAE | MAPE | RMSE |
|---|---|---|---|---|
| ARIMA-ETS | ARIMA | Single | Hybrid | Single |
|  | ETS | Hybrid | Single | Hybrid |
| ARIMA-NNAR | ARIMA | Hybrid | Hybrid | Hybrid |
|  | NNAR | Hybrid | Hybrid | Hybrid |
| ETS-NNAR | ETS | Hybrid | Hybrid | Hybrid |
|  | NNAR | Hybrid | Single | Hybrid |
| ARIMA-ETS-NNAR | ARIMA | Single | Hybrid | Hybrid |
|  | ETS | Hybrid | Hybrid | Hybrid |
|  | NNAR | Hybrid | Hybrid | Hybrid |

Table 8. Comparison of the efficiency between hybrid and single models.

| Hybrid | Single | Mild condition | | ICU | |
|---|---|---|---|---|---|
|  |  | MAE | RMSE | MAE | RMSE |
| ARIMA-NNAR | ARIMA | -5.15% | -5.21% | -1.87% | -4.19% |
|  | NNAR | -8.91% | -9.47% | -5.8% | -6.67% |
| ETS-NNAR | ETS | -6.15% | -6.53% | -9.02% | -12.4% |
|  | NNAR | -2.57% | -2.64% | -3.01% | -2.68% |
| ARIMA-ETS-NNAR | ARIMA | -0.69% | 0.15% | 0.39% | -0.1% |
|  | ETS | -8.13% | -8.3% | -9.61% | -12.4% |
|  | NNAR | -4.62% | -4.48% | -3.63% | -2.68% |

Notes: negative values show the percentage efficiency gains from using hybrid models.

Figure 2. Arima forecasts of patients hospitalized with mild symptoms and in ICU.

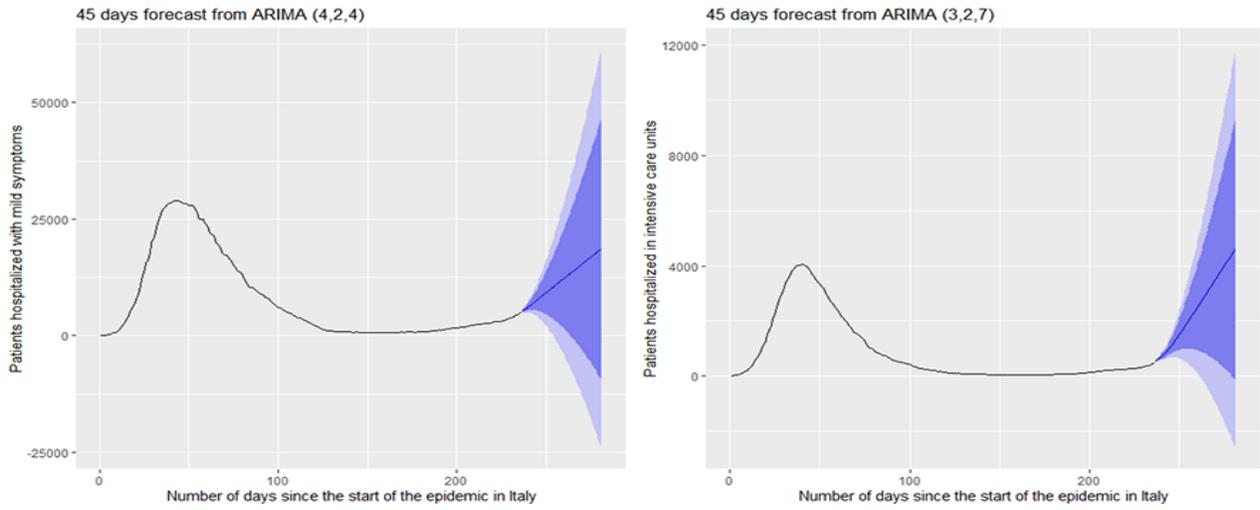

Figure 3. ETS forecasts of patients hospitalized with mild symptoms and in ICU.

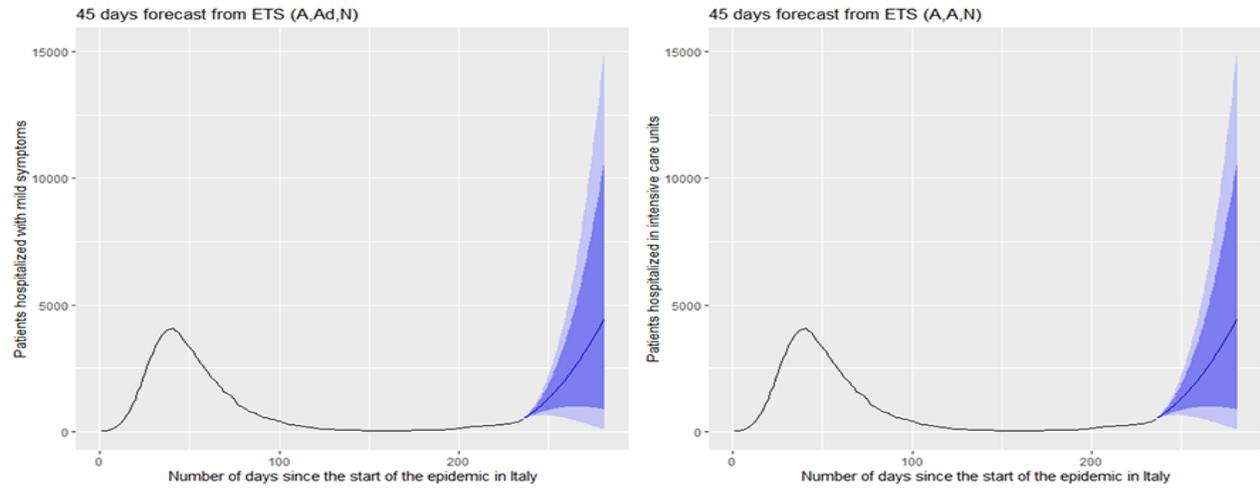

Figure 4. NNAR forecasts of patients hospitalized with mild symptoms and in ICU.

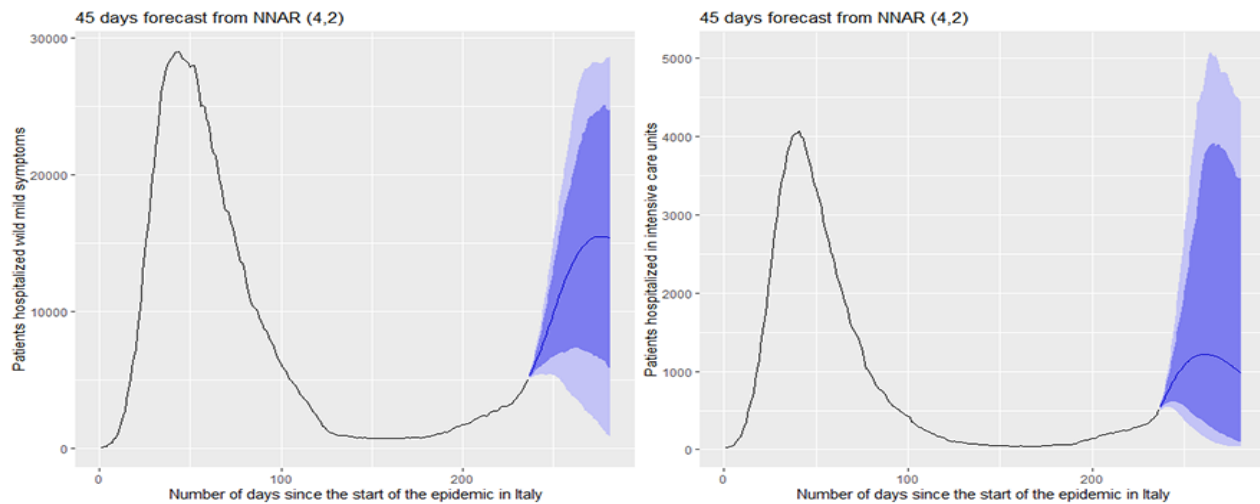

Figure 5. Hybrid ARIMA-ETS forecasts of patients hospitalized with mild symptoms and in ICU.

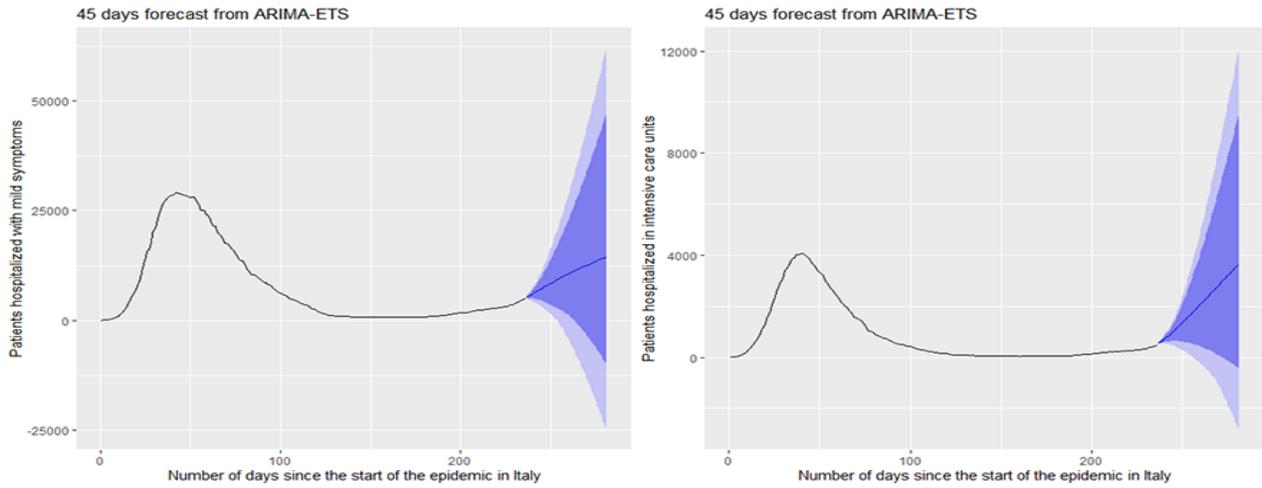

Figure 6. Hybrid ARIMA-NNAR forecasts of patients hospitalized with mild symptoms and in ICU.

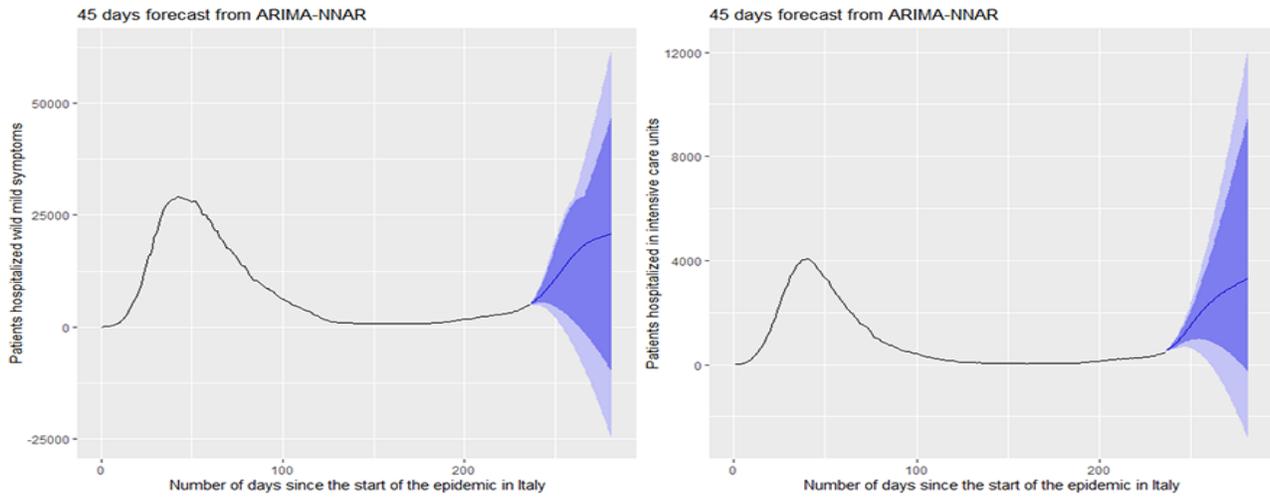

Figure 7. Hybrid ETS-NNAR forecasts of patients hospitalized with mild symptoms and in ICU.

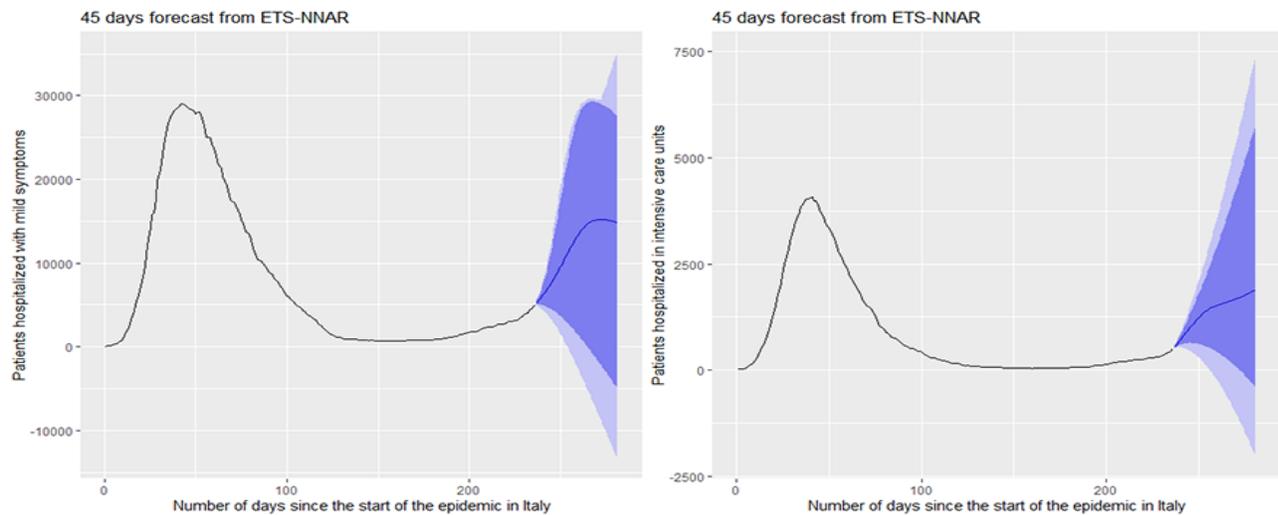

Figure 8. Hybrid ARIMA-ETS-NNAR forecasts of patients hospitalized with mild symptoms and in ICU.

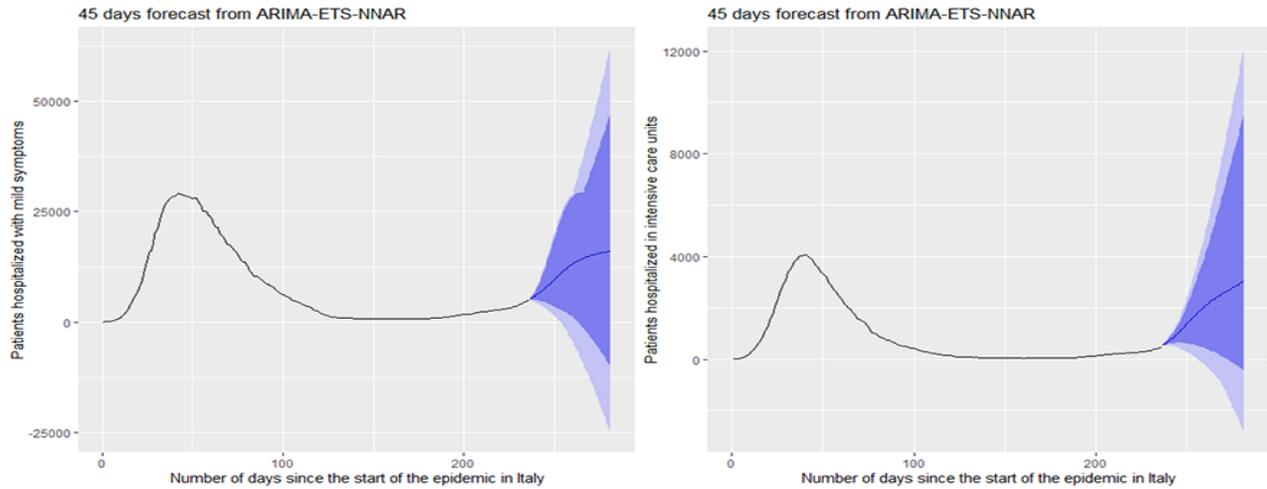

## 5. Conclusions

In this paper, I attempted to forecast the short and mid-term dynamics of the real-time patients hospitalized from COVID-19 in Italy. In particular, I used both single time series forecast methods and hybrid combinations of them. The results show that: i) the best single and hybrid models are ARIMA and ARIMA-NNAR, respectively; ii) and hybrid ARIMA-NNAR, ETS-NNAR, and ARIMA-ETS-NNAR models outperformed the respective single models, by leading to more accurate and reliable predictions. Thus, hybrid models seem to enhance the chances of capturing a greater number of combinations of the linear and nonlinear epidemic patterns, compared with the use of single time series forecasting methods.

Predictions seem also to give useful policy indications. In fact, they show that the number of patients hospitalized with mild symptoms and in ICU will significantly grow until mid-December 2020, when the second epidemic peak is expected. I predict that the necessary ordinary and intensive care beds will double in 10 days, and triple in about 20 days. Thus, it is necessary to strengthen the national health care system by buying protective equipment and hospital beds, managing health care facilities, and hiring and training healthcare workers.

Finally, the hybrid combination of ARIMA, ETS, and NNAR have proven to be sufficiently accurate in the short and mid-term. However, the results in the mid-term should be taken with more caution because of the inevitable uncertainty and bias, which tend to grow over time.

Sugiura, N. (1978). Further analysts of the data by akaike's information criterion and the finite corrections: Further analysts of the data by akaike's. *Communications in Statistics-Theory and Methods*, 7 (1), 13-26.

Swaraj A., Kaur A., Verma K., Singh G., Kumar A., & De Sales, L. M. (2020) Implementation of Stacking Based ARIMA Model for Prediction of Covid-19 Cases in India, 04 August 2020. Preprint (Version 1) available at Research Square.

Tuite, A. R., Ng, V., Rees, E., & Fisman, D. (2020). Estimation of COVID-19 outbreak size in Italy. *The Lancet infectious diseases*, 20 (5), 537.

Wieczorek, M., Siłka, J., & Woźniak, M. (2020). Neural network powered COVID-19 spread forecasting model. *Chaos, Solitons & Fractals*, 140, 110203.

Worldometer (2020). https://www.worldometers.info/coronavirus/.

Xu, C., Dong, Y., Yu, X., Wang, H., & Cai, Y. (2020). Estimation of reproduction numbers of COVID-19 in typical countries and epidemic trends under different prevention and control scenarios. *Frontiers of Medicine*, 1-10.

Yonar, H., Yonar, A., Tekindal, M. A., & Tekindal, M. (2020). Modeling and Forecasting for the number of cases of the COVID-19 pandemic with the Curve Estimation Models, the Box-Jenkins and Exponential Smoothing Methods. *EJMO*, 4 (2), 160-165.

Zhang, G. P. (2003). Time series forecasting using a hybrid ARIMA and neural network model. *Neurocomputing*, 50, 159-175.

Zhao, S., Lin, Q., Ran, J., Musa, S. S., Yang, G., Wan, W., Lou, Y., Gao, D., Yang, L., He, D., & Wang, M. H. (2020). Preliminary estimation of the basic reproduction number of novel coronavirus (2019-nCoV) in China, from 2019 to 2020: A data-driven analysis in the early phase of the outbreak. *International Journal of Infectious Diseases*, 92, 214-217.

Zhou, T., Liu, Q., Yang, Z., Liao, J., Yang, K., Bai, W., Xin, L., & Zhang, W. (2020). Preliminary prediction of the basic reproduction number of the Wuhan novel coronavirus 2019-nCoV. *Journal of Evidence-Based Medicine*, 13 (1), 2-7.

Appendix A.

Table A1. Estimated parameters for ARIMA models (Fig. 2).

| Parameters | Mild condition | | ICU | |
|---|---|---|---|---|
| | Coefficients | Standard error | Coefficients | Standard error |
| AR (1) | -0.6491*** | 0.1171 | -0.0924 | 0.1715 |
| AR (2) | 0.7842*** | 0.1603 | 0.0094 | 0.1364 |
| AR (3) | 0.3078*** | 0.1084 | 0.5872*** | 0.1314 |
| AR (4) | -0.3912*** | 0.0772 | | |
| MA (1) | 0.1763* | 0.0984 | -0.4972*** | 0.171 |
| MA (2) | -1.3753*** | 0.1034 | -0.1028 | 0.2107 |
| MA (3) | 0.003 | 0.0689 | -0.4019** | 0.1697 |
| MA (4) | 0.803*** | 0.0762 | 0.3012** | 0.1225 |
| MA (5) | | | 0.0347 | 0.0794 |
| MA (6) | | | -0.0566 | 0.0787 |
| MA (7) | | | 0.3411*** | 0.0856 |

Notes: ***p-value < 0.01; **p-value < 0.05; *p-value < 0.1.

Table A2. Estimated parameters for ETS models (Fig.3).

| Smoothing parameters | Mild condition | ICU |
|---|---|---|
| | Coefficients | Coefficients |
| $\alpha$ | 0.9999 | 0.8734 |
| $\beta$ | 0.3635 | 0.4243 |
| $\rho$ | 0.98 | |

Table A3. Estimated parameters for ARIMA models (Fig. 5, 6, & 8)

| Parameters | Mild condition | | ICU | |
|---|---|---|---|---|
| | Coefficients | Standard error | Coefficients | Standard error |
| AR (1) | -0.6491*** | 0.1171 | 0.3273*** | 0.0819 |
| AR (2) | 0.7842*** | 0.1603 | 0.8596*** | 0.0877 |
| AR (3) | 0.3078*** | 0.1084 | -0.1893** | 0.0776 |
| AR (4) | -0.3912*** | 0.0772 | -0.1597** | 0.0717 |
| MA (1) | 0.1763* | 0.0984 | -0.9059*** | 0.0517 |
| MA (2) | -1.3753*** | 0.1034 | -0.756*** | 0.0754 |
| MA (3) | 0.003 | 0.0689 | 0.8659*** | 0.0464 |
| MA (4) | 0.803*** | 0.0762 | | |

Notes: ***p-value < 0.01; **p-value < 0.05; *p-value < 0.1.

Table A4. Estimated parameters for ETS models (Fig. 5, 7, & 8)

| Smoothing parameters | Mild condition | ICU |
|---|---|---|
| | Coefficients | Coefficients |
| $\alpha$ | 0.9999 | 0.9071 |
| $\beta$ | 0.4555 | 0.5871 |
| $\rho$ | 0.9708 | |

Table A5. The predicted values of hospitalized with mild condition and in ICU in Italy, from October 14, 2020 to November 27, 2020.

| | Hospitalized with mild condition* | | | Hospitalized in intensive care units* | | |
|---|---|---|---|---|---|---|
| Date | ARIMA-NNAR | ETS-NNAR | ARIMA-ETS-NNAR | ARIMA-NNAR | ETS-NNAR | ARIMA-ETS-NNAR |
| 14-10-2020 | 5,341 | 5,325 | 5,327 | 564 | 561 | 562 |
| 15-10-2020 | 5,643 | 5,583 | 5,595 | 607 | 608 | 607 |
| 16-10-2020 | 5,957 | 5,847 | 5,867 | 664 | 660 | 661 |
| 17-10-2020 | 6,307 | 6,117 | 6,157 | 720 | 713 | 713 |
| 18-10-2020 | 6,672 | 6,403 | 6,453 | 786 | 766 | 772 |
| 19-10-2020 | 7,065 | 6,697 | 6,765 | 853 | 820 | 832 |
| 20-10-2020 | 7,472 | 7,006 | 7,084 | 929 | 875 | 896 |
| 21-10-2020 | 7,906 | 7,331 | 7,419 | 1,006 | 931 | 961 |
| 22-10-2020 | 8,359 | 7,673 | 7,767 | 1,091 | 986 | 1,030 |
| 23-10-2020 | 8,829 | 8,032 | 8,125 | 1,177 | 1,040 | 1,100 |

| Date | | | | | | |
|---|---|---|---|---|---|---|
| 24-10-2020 | 9,322 | 8,407 | 8,496 | 1,266 | 1,093 | 1,173 |
| 25-10-2020 | 9,817 | 8,793 | 8,868 | 1,357 | 1,145 | 1,246 |
| 26-10-2020 | 10,333 | 9,189 | 9,251 | 1,449 | 1,193 | 1,320 |
| 27-10-2020 | 10,847 | 9,593 | 9,631 | 1,541 | 1,240 | 1,394 |
| 28-10-2020 | 11,377 | 10,003 | 10,017 | 1,633 | 1,283 | 1,469 |
| 29-10-2020 | 11,905 | 10,416 | 10,398 | 1,723 | 1,323 | 1,542 |
| 30-10-2020 | 12,442 | 10,832 | 10,778 | 1,813 | 1,359 | 1,616 |
| 31-10-2020 | 12,977 | 11,245 | 11,152 | 1,899 | 1,392 | 1,688 |
| 1-11-2020 | 13,510 | 11,655 | 11,518 | 1,984 | 1,420 | 1,759 |
| 2-11-2020 | 14,038 | 12,058 | 11,877 | 2065 | 1,446 | 1,829 |
| 3-11-2020 | 14,553 | 12,450 | 12,220 | 2,143 | 1,468 | 1,896 |
| 4-11-2020 | 15,063 | 12,826 | 12,552 | 2,218 | 1,487 | 1,962 |
| 5-11-2020 | 15,550 | 13,182 | 12,862 | 2,290 | 1,505 | 2,025 |
| 6-11-2020 | 16,025 | 13,514 | 13,155 | 2,358 | 1,521 | 2,086 |
| 7-11-2020 | 16,470 | 13,818 | 13,425 | 2,423 | 1,535 | 2,144 |
| 8-11-2020 | 16,894 | 14,092 | 13,676 | 2,485 | 1,550 | 2,200 |
| 9-11-2020 | 17,287 | 14,332 | 13,904 | 2,543 | 1,564 | 2,254 |
| 10-11-2020 | 17,651 | 14,540 | 14,114 | 2,599 | 1,578 | 2,305 |
| 11-11-2020 | 17,988 | 14,715 | 14,306 | 2,652 | 1,592 | 2,354 |
| 12-11-2020 | 18,292 | 14,859 | 14,481 | 2,702 | 1,606 | 2,401 |
| 13-11-2020 | 18,573 | 14,974 | 14,643 | 2,750 | 1,622 | 2,446 |
| 14-11-2020 | 18,824 | 15,062 | 14,789 | 2,796 | 1,638 | 2,490 |
| 15-11-2020 | 19,057 | 15,127 | 14,927 | 2,841 | 1,654 | 2,533 |
| 16-11-2020 | 19,264 | 15,170 | 15,051 | 2,883 | 1,672 | 2,575 |
| 17-11-2020 | 19,458 | 15,194 | 15,169 | 2,925 | 1,690 | 2,617 |
| 18-11-2020 | 19,632 | 15,202 | 15,277 | 2,966 | 1,709 | 2,658 |
| 19-11-2020 | 19,795 | 15,196 | 15,380 | 3,006 | 1,729 | 2,698 |
| 20-11-2020 | 19,945 | 15,177 | 15,476 | 3,045 | 1,749 | 2,739 |
| 21-11-2020 | 20,085 | 15,148 | 15,568 | 3,084 | 1,770 | 2,780 |
| 22-11-2020 | 20,218 | 15,109 | 15,656 | 3,123 | 1,791 | 2,821 |
| 23-11-2020 | 20,342 | 15,063 | 15,739 | 3,162 | 1,813 | 2,862 |
| 24-11-2020 | 20,463 | 15,010 | 15,821 | 3,201 | 1,836 | 2,903 |
| 25-11-2020 | 20,576 | 14,952 | 15,900 | 3,240 | 1,859 | 2,945 |
| 26-11-2020 | 20,688 | 14,889 | 15,977 | 3,280 | 1,883 | 2,988 |
| 27-11-2020 | 20,795 | 14,823 | 16,052 | 3,320 | 1,907 | 3,030 |

*Values are rounded to the nearest integer.